\documentstyle[aps,prb,eqsecnum,epsf,floats]{revtex}
\begin{document}
% Macros for the various macro package names, etc.
\def\SNG{{\em Physical Review Style and Notation Guide}}
\def\LUG {{\em \LaTeX{} User's Guide \& Reference Manual}}
\def\btt#1{{\tt$\backslash$\string#1}}%
\def\REVTeX{REV\TeX}
\def\AmS{{\protect\the\textfont2
        A\kern-.1667em\lower.5ex\hbox{M}\kern-.125emS}}
\def\AmSLaTeX{\AmS-\LaTeX}
\def\BibTeX{\rm B{\sc ib}\TeX}
%\makeatletter
%\tighten
\twocolumn[\hsize\textwidth\columnwidth\hsize\csname@twocolumnfalse%
\endcsname

\title{Quantum critical behavior of clean itinerant 
       ferromagnets$^*$\\
       \small{$[$ Z. Phys. B {\bf 103}, 451 (1997) $]$}}

\author{Thomas Vojta}
\address{Department of Physics and Materials Science Institute,
         University of Oregon, Eugene, OR 97403\\
         and
         Institut f{\"u}r Physik, TU Chemnitz-Zwickau, D-09107 Chemnitz, FRG}
\author{D.Belitz and R.Narayanan}
\address{Department of Physics and Materials Science Institute,
         University of Oregon, Eugene, OR 97403}
\author{T.R.Kirkpatrick}
\address{Institute for Physical Science and Technology, and Department of 
         Physics\\
         University of Maryland, College Park, MD 20742}

\date{\today}
\maketitle

\begin{abstract}
We consider the quantum ferromagnetic transition at zero temperature 
in clean itinerant electron systems. We find that the Landau-Ginzburg-Wilson
order parameter field theory breaks down since the electron-electron
interaction leads to singular coupling constants in the 
Landau-Ginzburg-Wilson functional. These couplings generate an effective
long-range interaction between the spin or order parameter fluctuations
of the form $1/r^{2d-1}$, with $d$ the spatial dimension. This leads to
unusual scaling behavior at the quantum critical point in $1<d\leq 3$, 
which we determine exactly. We also discuss the quantum-to-classical 
crossover at small but finite temperatures, which is characterized by
the appearance of multiple temperature scales. A comparison with
recent results on disordered itinerant ferromagnets is given.
\end{abstract}

\pacs{PACS numbers: 64.60.Ak , 75.10.Jm , 75.40.Cx , 75.40.Gb}
]
%\narrowtext
\section{Introduction}
\label{sec:I}

\footnotetext{$^*$Dedicated to Prof. W. G{\"o}tze on the occasion of his
sixtieth birthday}
The physics of quantum phase transitions has been a subject of
great interest lately. In contrast to the ususal classical or thermal phase
transitions, quantum phase transitions occur at zero temperature as
a function of some nonthermal control parameter, and the fluctuations
that drive the transition are
of quantum nature rather than thermal in origin. Among the 
transitions that have been
investigated are various metal-insulator transitions, the 
superconductor-insulator transition in thin metal
films, and a variety of magnetic 
phase transitions. Early work in this field\cite{Hertz} established that 
if the quantum phase transition has a
classical analog at finite temperature, then the former tends to have a
simpler critical behavior in the physical dimensions $d=3$ or $d=2$ 
than the latter. The reason for this tendency is 
that the coupling between statics and dynamics that is inherent to quantum
statistics problems effectively increases the dimensionality of the system
from $d$ to $d+z$, with $z$ the dynamical critical exponent. 
This reduces the upper critical dimension $d_c^+$, which is the 
dimension above which mean-field theory yields the exact critical behavior,
by $z$ from its value for the classical transition.

One of the most obvious examples of a quantum phase transition, and the
first one studied in detail, is the ferromagnetic transition of itinerant
electrons that occurs as a function of the 
exchange coupling between the electron spins. In a pioneering paper,
Hertz\cite{Hertz} derived a Landau-Ginzburg-Wilson (LGW) functional for 
this transition by considering a simple model of itinerant electrons
that interact only via the exchange interaction in the particle-hole
spin-triplet channel. Hertz analyzed this LGW functional by means of 
renormalization group (RG) methods that generalize Wilson's treatment of 
classical phase transitions. He concluded that the critical behavior in the 
physical dimensions $d=3$ and $d=2$ is mean-field like,
since the dynamical critical exponent $z=3$ decreases the upper critical
dimension from $d_c^+ =4$ for the classical case to $d_c^+=1$ in the
quantum case. In order to study nontrivial quantum critical behavior,
Hertz then considered a model with a magnetization that is defined
on a space of arbitrary dimension $d$, while the correlation functions that 
determine the coefficients in the LGW functional are taken to be those of a 
$3$-$d$ Fermi gas. For this model, he calculated non mean-field like
critical behavior in $d<1$ by means of a $1-\epsilon$ expansion.
Despite the somewhat artificial
nature of this model, there is a general belief that the qualitative
features of Hertz's analysis, in particular the fact that there is mean-field
like critical behavior for all $d>1$, apply to real itinerant
quantum ferromagnets as well.

In this paper we reexamine the ferromagnetic quantum phase transition of 
itinerant electrons and show that the above belief is 
qualitatively mistaken. We first consider a model that is more realistic
than Hertz's, viz. with an electron-electron interaction that is not 
restricted to the particle-hole spin-triplet channel. We find that
the LGW approach breaks down due to the
presence of soft modes in addition to the critical modes, namely 
particle-hole excitations that couple to the conserved order parameter.
These soft modes are integrated
out in the derivation of the LGW functional, and this
leads to singular vertices in the order parameter field theory.
This is a rather general observation; analogous effects are expected
for a large class of quantum phase transitions, and in general they
invalidate the application to quantum phase transitions of the classical 
LGW philosophy of deriving an effective local field theory entirely in 
terms of the order parameter.\cite{LGWfootnote} However, for the present
problem we find that the mathematical difficulties generated by
the singular couplings can be handled.
The resulting nonlocal field theory contains an
effective long-range interaction between the order parameter fluctuations,
and we are able to determine the critical behavior
exactly for all $d>1$. For $1<d\leq 3$ the result is different from either 
mean-field critical behavior or classical Heisenberg critical behavior.
We then show that the same is true for Hertz's
original model. Our results invalidate the $1-\epsilon$
expansion in that paper, and it corroborates and explains an observation 
made recently
by Sachdev,\cite{Sachdev} who noted that Hertz's results in $d<1$ cannot
be correct since they violate an exact exponent equality.

The outline of this paper is as follows. In Sec.\ \ref{sec:II} we first
define our model of itinerant electrons, and then discuss how to
derive an order parameter description for the ferromagnetic phase transition
starting from a fermionic field theory. We discuss the behavior of 
the coefficients in the resulting LGW functional and show that they
have long-range properties. In Sec.\ \ref{sec:III} the quantum 
critical behavior of the resulting nonlocal field theory at zero temperature
is determined exactly for dimensions $d>1$.
For $d>3$, mean-field exponents are obtained, while for $1<d<3$, 
$d$-dependent exponents are found. In Sec.\ \ref{sec:IV} we investigate
the behavior at small but finite temperatures 
which is characterized by multiple independent temperature scales.
In Sec.\ \ref{sec:V} we reexamine Hertz's original model of the ferromagnetic
quantum phase transition. We show that upon renormalization it acquires the
same features as the more realistic model studied in 
Secs.\ \ref{sec:II} - \ref{sec:IV}.
In Sec.\ \ref{sec:VI} we discuss our results,
and in particular the relation of the present paper
to recent work on {\em disordered} itinerant quantum magnets.\cite{fm}
A technical detail concerning perturbation theory is relegated to 
Appendix\ \ref{app:A}, while in
Appendix \ \ref{app:B} we discuss logarithmic
corrections to scaling that exist in $d=3$ as well as for all dimensions
$1<d<3$. A short account of some of our results has previously been 
published.\cite{Letter}

\section{Order parameter field theory for the quantum ferromagnetic transition}
\label{sec:II}

In the first part of this section we define our model of itinerant interacting
electrons in terms of a fermionic field theory. Since we are interested in
a phase transition to a magnetically ordered phase, we choose the spin
density fluctuation as our order parameter, and proceed to
derive an LGW or order parameter description of this transition.
We then derive and discuss the coefficients in this LGW functional.
As we will see,
the interactions in the effective LGW theory are long-ranged due to soft
modes that have been integrated out in deriving the LGW functional.

\subsection{The Model}
\label{subsec:II.A}

The partition function of any fermionic system can be written in the 
form,\cite{NegeleOrland}
\begin{mathletters}
\label{eqs:2.1}
\begin{equation}
Z=\int D{\bar\psi}\,D\psi\ \exp\left(S\left[{\bar\psi},\psi\right]\right)\quad,
\label{eq:2.1a}
\end{equation}
where $\bar\psi$ and $\psi$ are Grassmannian (i.e., anticommuting) fields.
$D{\bar\psi}\,D\psi$ denotes the Grassmannian functional integration measure,
and $S$ is the action,
\begin{equation}
S = \int_0^{\beta} d\tau \int d{\bf x}\ 
   {\bar\psi}^i({\bf x},\tau)\,{\partial\over\partial\tau}\,\psi_i({\bf x},\tau)
 - \int_0^{\beta} d\tau\ H(\tau)\quad.
\label{eq:2.1b}
\end{equation}
\end{mathletters}%
Here ${\bf x}$ denotes position and $\tau$ imaginary time, $H(\tau)$ 
is the Hamiltonian in imaginary time representation,
$\beta=1/T$ is the inverse temperature, $i=1,2$ denotes spin labels, and
summation over repeated covariant and contravariant spin indices is implied. 
Throughout this paper we use
units such that $k_B = \hbar = e^2 = 1$. We start with a microscopic model 
of itinerant, interacting electrons,
\begin{eqnarray}
H(\tau) = \int d{\bf x}\ {\bar\psi}^i({\bf x},\tau) \left[ 
          - { 1\over 2m}\,\nabla^2 - \mu \right] \psi_i({\bf x}, \tau)
\nonumber\\
          +\ {1\over 2} \int d{\bf x}\,d{\bf y}\ u({\bf x}-{\bf y})\ 
             {\bar\psi}^i({\bf x},\tau)\,{\bar\psi}^j({\bf y},\tau)\,
             \psi_j({\bf y},\tau)\,
\nonumber\\
             \times\psi_i({\bf x},\tau)\quad.
\label{eq:2.2}
\end{eqnarray}
Here $m$ is the electron mass, $\mu$ is the chemical potential, and
$u({\bf x}-{\bf y})$ is the electron-electron interaction potential.
More realistic models to describe itinerant electron
magnetism including, e.g., band structure, can be considered along the same
lines. The salient points of the present paper, however,
are due to long-wavelength effects and therefore
are independent of microscopic details like the band structure.
For our purposes it therefore is sufficient to study the model defined in
Eq.\ (\ref{eq:2.2}).

In order to describe magnetism, it is convenient and standard practice to
break the interaction part of the action $S$, which we denote by $S_{int}$,
into particle-hole spin-singlet, particle-hole spin-triplet and 
particle-particle or Cooper channel contributions, which we denote by 
$S_{int}^{\,(s)}$, 
$S_{int}^{\,(t)}$ and $S_{int}^{\,(c)}$, respectively.\cite{AGD} 
For simplicity, we assume that the interactions are short-ranged in all of
these channels. In a metallic system this is justified due to screening,
and an effective model with a short-ranged interactions can be derived 
starting from a bare Coulomb interaction.\cite{AGD} 
The spin-triplet interaction warrants special attention, since the 
interactions between spin density fluctuations that are described by
$S_{int}^{\,(t)}$ are what causes ferromagnetism. 
We therefore consider this part of the action separately, and write
\begin{equation}
S = S_0 + S_{int}^{\,(t)}\quad,
\label{eq:2.3}
\end{equation}
with
\begin{mathletters}
\label{eqs:2.4}
\begin{equation}
S_{int}^{\,(t)} = {\Gamma_t\over 2} \int d{\bf x}\,d\tau\ {\bf n}_s({\bf x},\tau)
                 \cdot {\bf n}_s({\bf x},\tau)\quad,
\label{eq:2.4a}
\end{equation}
where ${\bf n}_s$ is the electron spin density vector with components,
\begin{equation}
n_s^a({\bf x},\tau) = {1\over 2}\ {\bar\psi}^i({\bf x},\tau)\,
     \sigma^a_{ij}\,\psi^j({\bf x},\tau)\quad.
\label{eq:2.4b}
\end{equation}
\end{mathletters}%
Here the $\sigma^a$ ($a=1,2,3$) are the Pauli matrices, and $\Gamma_t$ is the
spin-triplet interaction amplitude.
$S_0$ in Eq.\ (\ref{eq:2.3}) contains the free electron part and all 
interaction parts other than the particle-hole spin-triplet contribution
to the action. It reads explicitly, 
\begin{mathletters}
\label{eqs:2.5}
\begin{eqnarray}
S_0 = \int_0^{\beta} d\tau \int d{\bf x}\ \biggl[
   {\bar\psi}^i({\bf x},\tau)\,{\partial\over\partial\tau}\,\psi_i({\bf x},\tau)
\nonumber\\
    - {\bar\psi}^i({\bf x},\tau) \left[ 
          - \nabla^2/2m - \mu \right] \psi_i({\bf x}, \tau) \biggr]
\nonumber\\
   - {\Gamma_s\over 2} \int_0^{\beta} d\tau \int d{\bf x}\ 
         n({\bf x},\tau)\,n({\bf x},\tau)\quad
\nonumber\\
   - \Gamma_c \int_0^{\beta} d\tau \int d{\bf x}\
         {\bar n}_c^{ij}({\bf x},\tau)\,n_{c,ij}({\bf x},\tau)\quad.
\label{eq:2.5a}
\end{eqnarray}
Here $n$ is the electron number or charge density,
\begin{equation}
n({\bf x},\tau) = {\bar\psi}^i({\bf x},\tau)\, \psi_i({\bf x},\tau)\quad,
\label{eq:2.5b}
\end{equation}
and $n_{c,ij}$ is the anomalous or Cooper channel density,
\begin{equation}
n_{c,ij}({\bf x},\tau) = \left(1 - \delta_{ij}\right)\,\psi_i({\bf x},\tau)\, 
                          \psi_j({\bf x},\tau)\quad,
\label{eq:2.5c}
\end{equation}
\end{mathletters}%
(no summation). To avoid double counting, only the long-wavelength
fluctuations of the charge density and the anomalous density must be
taken into account in Eq.\ (\ref{eq:2.5a}).\cite{decompositionfootnote}
$\Gamma_s$ and $\Gamma_c$ are the interaction
amplitudes in the particle-hole spin-singlet and in the Cooper channel,
respectively. For our purposes it is sufficient to treat $\Gamma_t$, 
$\Gamma_s$, and $\Gamma_c$ simply as numbers, although in a more complete
model they are complicated short-ranged interaction 
potentials.\cite{AGD,Landaufootnote}
For later reference we note that they result from averaging the original
interaction over different regions of phase space, so that the three
interaction amplitudes are independent parameters.

\subsection{Order parameter field theory}
\label{subsec:II.B}

The standard theory of continuous {\em thermal} phase 
transitions proceeds from a
microscopic model by identifying the order parameter relevant for the
transition, and by integrating out all degrees of freedom except for
the long wavelength order parameter fluctuations,
or critical modes. The result of this procedure is a LGW theory, i.e.
an effective field theory for the critical modes.\cite{WilsonKogut}
The rationale behind this approach is that these fluctuations, 
which are slowly varying 
in space, determine the behavior near the critical point. The same basic
idea has been applied to {\em quantum} phase transitions, with the only
principal difference being that the critical modes are now slowly varying 
in both space and (imaginary) time. While we will use
this approach here, motivated in part by previous work on itinerant
electronic systems,\cite{Hertz} we mention that in general one
should worry about both the critical modes, and all other slow or soft 
modes, even if these other soft modes are not `critical' in the sense
that they change their character at the phase transition. Of course
this concern is not restricted to quantum phase transitions; additional
soft modes can and do occur also at thermal phase transitions. However,
while this is an exceptional occurence in the case of the latter, we will
argue below that it poses a more serious problem for the former,
since at $T=0$ there are more soft modes than at finite temperature.
We will see that in the present problem, and in a large class of other
quantum phase transitions, such additional modes are 
indeed present and lead to complications if one insists on a description
entirely in terms of the order parameter. For the
problem under consideration, however, we will be able to overcome
these problems and determine the critical behavior within the framework
of an LGW theory.

The techniques for deriving an order parameter field theory, starting
with Eqs.\ (\ref{eqs:2.1}) through (\ref{eqs:2.5}), are standard.\cite{Hertz} 
We introduce a classical vector field ${\bf M}({\bf x},\tau)$ with
components $M^a$, the average of one of which ($M^3$, say) is
proportional to the magnetization $m$, and decouple the four fermion term in 
$S_{int}^{\,(t)}$ by performing a Hubbard-Stratonovich
transformation. Subsequently, all degrees of freedom other than ${\bf M}$ are
integrated out. This procedure in particular integrates out soft
particle-hole excitations, which are the additional soft modes mentioned
above. We obtain the partition function $Z$ in the form,
\begin{mathletters}
\label{eqs:2.6}
\begin{equation}
Z = e^{-F_0/T} \int D[{\bf M}]\ \exp\left(-\Phi[{\bf M}]\right)\quad,
\label{eq:2.6a}
\end{equation}
with $F_0$ the noncritical part of the free energy. The LGW functional
$\Phi$ reads,
\begin{eqnarray}
\Phi[&&{\bf M}] = {\Gamma_t\over 2} \int dx\ {\bf M}(x)\cdot {\bf M}(x)
                \quad\quad\quad\quad\quad\quad
\nonumber\\
              &&- \ln \left<\exp\left[-\Gamma_t \int dx\ {\bf M}(x)\cdot
                              {\bf n}_s(x)\right]\right>_{S_0}\quad.
\label{eq:2.6b}
\end{eqnarray}
\end{mathletters}%
Here we use a four-vector notation with
$x = ({\bf x},\tau)$, and $\int dx = \int d{\bf x} \int_0^{\beta} d\tau$,
and $\langle\ldots\rangle_{S_0}$ denotes an average taken with the
action $S_0$. A formal expansion of $\Phi$ in powers of ${\bf M}$ takes
the form,
\begin{mathletters}
\label{eqs:2.7}
\begin{eqnarray}
\Phi[{\bf M}] = {1\over 2} \int dx\,dy\ \left[{1\over \Gamma_t}\delta(x-y) 
                                                      - \chi^{(2)}(x-y)\right]
\nonumber\\
                                        \times {\bf M}(x)\cdot {\bf M}(y)
\nonumber\\
              + \sum_{n=3}^{\infty}b_n \int dx_1\,\ldots\,dx_n\
\nonumber\\
                \times \chi^{(n)}_{a_1\ldots a_n} (x_1-x_n,\ldots,x_{n-1}-x_n) 
\nonumber\\
               \times M^{a_1}(x_1)\,\ldots\,M^{a_n}(x_n)\quad ,
\label{eq:2.7a}
\end{eqnarray}
where $b_n = (-1)^n/n!$ and we have scaled ${\bf M}$ with $\Gamma_t^{-1}$. 
The coefficients $\chi^{(n)}$ in Eq.\ (\ref{eq:2.7a}) are connected
n-point spin density correlation functions of a
reference system whose action is given by
$S_0$. They are defined as,
\begin{equation}
\chi_{a_1\ldots a_n}^{\,(n)}(x_1,\ldots,x_n) = \langle n_s^{a_1}(x_1)\cdots
                       n_s^{a_n}(x_n)\rangle^c_{S_0}\quad,
\label{eq:2.7b}
\end{equation}
\end{mathletters}%
where the superscript `c' denotes a cumulant or connected correlation
function. The tensor structure of $\chi^{(n)}$ is restricted by rotational
invariance in spin space. For instance, $\chi^{(3)}$ is proportional to the
completely antisymmetric third-rank tensor $\epsilon^{abc}$. 
For our simple model, the reference ensemble $S_0$
consists of free electrons with short-ranged particle-hole spin-singlet
and particle-particle model interactions. As mentioned
in Sec.\ \ref{subsec:II.A} above, the model can be made more realistic, if
desired, by e.g. including band structure effects. To this end
one would simply replace the $\chi^{(n)}$ above with correlation functions
for band electrons.

\subsection{The coefficients of the LGW functional}
\label{subsec:II.C}

To proceed, we have to calculate the coefficients in the LGW functional,
Eq.\ (\ref{eq:2.7a}), i.e. the connected spin density correlation functions 
of the reference system $S_0$. Normally, one would localize the
individual terms in Eq.\ (\ref{eq:2.7a}) about a single point in space and
time, and expand the correlation functions in powers of gradients and
frequencies. However, for the system under consideration such a gradient 
expansion does not exist. To illustrate this point,
let us first consider the spin susceptibility
$\chi^{(2)}_{ab} \equiv \delta_{ab}\,\chi_s$,
whose Fourier transform is defined by 
\begin{eqnarray}
\chi_s({\bf q},\Omega_n) = \int d(&&{\bf x}_1-{\bf x}_2)\,d(\tau_1-\tau_2)\ 
            e^{-i{\bf q}\cdot ({\bf x}_1-{\bf x}_2)}
\nonumber\\
                                     &&\times e^{i\Omega_n (\tau_1-\tau_2)} 
            \ \chi_s(x_1-x_2)\quad.
\label{eq:2.8}
\end{eqnarray}
where $\Omega_n = 2\pi Tn$ is a bosonic Matsubara frequency. 
In order to study the critical
behavior we have to determine $\chi^{(2)}$ in the long-wavelength 
and low-frequency limit. In a system with a conserved order parameter,
the frequency must be taken to zero {\em before} the wavenumber, otherwise
one never reaches criticality.\cite{limitsfootnote}
In the critical region we therefore have $\vert\Omega_n\vert << q$. 
In this limit the spin susceptibility has the structure
\begin{equation}
\chi_s({\bf q},\Omega_n) = \chi_0(q)\ \left[ 1 -
      \vert\Omega_n\vert/q + \ldots\right]\quad,
\label{eq:2.9}
\end{equation}
where $q$ and $\Omega_n$ are being measured in suitable units, and 
$\chi_0(q)$ is the static spin susceptibility of the reference system.

The static spin susceptibility of non-interacting electrons, i.e.
the Lindhard function, as a function of the wavenumber $q$ is analytic 
at $q=0$. However, for any electron system with a nonvanishing 
particle-hole spin-triplet interaction, $\Gamma_t \neq 0$, 
there is a nonanalytic correction to the static spin
susceptibility.\cite{chi_s}
Although our bare reference system $S_0$ does not contain such 
an interaction amplitude, in a generic reference system a nonvanishing
$\Gamma_t$ is generated in perturbation theory.
For being generic in this sense, it suffices that the bare
system has a nonvanishing Cooper channel interaction amplitude 
$\Gamma_c \neq 0$. This is shown in Appendix \ref{app:A}. 
Effectively, we therefore have 
to calculate the static spin susceptibility of a paramagnetic Fermi 
liquid with a nonvanishing $\Gamma_t$. It has been shown in
Ref.\ \onlinecite{chi_s} that the result is,
\begin{mathletters}
\label{eqs:2.10}
\begin{equation}
\chi_0(q\rightarrow 0) = c_0 - c_{d-1}\,q^{d-1} - c_2\,q^2 + \ldots\quad,
\label{eq:2.10a}
\end{equation}
where $c_2>0$, and $c_{d-1}$ is proportional to $\Gamma_t^2$ for small
$\Gamma_t$.\cite{signfootnote} The nontrivial, and for our purposes
most interesting, contribution in Eq.\ (\ref{eq:2.10a}) is the nonanalytic
term $\sim q^{d-1}$. The above form holds for
$1<d<3$. In $d=3$ the nonanalyticity is of the form $q^2 \ln q$,
and for $d>3$ the leading $q$-dependence is given by the quadratic term.

This feature of a Fermi liquid is of crucial importance for what follows.
It therefore warrants some discussion, even though it has recently been
discussed in great detail.\cite{chi_s} It is well known that in a Fermi
liquid, both in $d=3$ and in $d=2$, the specific heat and the quasiparticle
lifetime are nonanalytic functions of the temperature.\cite{BaymPethick}
The physical reason for these effects is, in the language of the present
paper, the soft particle-hole excitations that are always present in a
electron system at zero temperature. Although they become massive at nonzero
temperature (see Eq.\ (\ref{eq:2.10b}) below), 
this is sufficient to make various observables
nonanalytic functions of temperature at $T=0$. The nonanalytic wavenumber
dependence of the spin susceptibility at $T=0$ is just another manifestation
of this effect. It is analogous to a feature of {\em disordered}
electron systems, for which it is known that $\chi_0(q\rightarrow 0)
\sim {\rm const.} + q^{d-2}$.\cite{fm,ernst} Since the physical reason for this
effect is again the presence of soft modes, it is not qualitatively tied 
to disordered systems. The only difference between disordered and clean
systems in this respect is the nature of the soft modes: In the former
they are diffusive, leading to an exponent $d-2$, while in the latter they
are ballistic and lead to an exponent $d-1$.

In the present context, the nonanalytic behavior of $\chi_0$ implies that 
the standard gradient
expansion mentioned at the beginning of this subsection does not exist.
Rather, the interaction between the order parameter fluctuations is
effectively of long range, and 
in real space takes the form $r^{-(2d-1)}$. This is true only at $T=0$.
At finite temperature,
where one has to perform a frequency sum rather than a frequency integral
to calculate the correlation function,
the nonanalytic term is replaced by a term of the schematic 
structure,\cite{chi_s}
\begin{equation}
q^{d-1} \rightarrow (q + T)^{(d-1)}\quad,
\label{eq:2.10b}
\end{equation}
\end{mathletters}%
so for fixed $T>0$ an analytic expansion about $q=0$ exists, and the standard
local LGW functional is obtained.

We now turn to the higher spin density correlation functions
$\chi^{(n)}$. The same physics that causes the nonanalyticity
in the spin susceptibility, Eq.\ (\ref{eq:2.10a}), leads to an even
stronger effect in the higher
correlation functions, and results in $\chi^{(n)}$ for $d<n-1$ not being 
finite in the limit of zero
frequencies and wave numbers. One finds
that in this limit $\chi^{(n)}$ is schematically given by
\begin{mathletters}
\label{eqs:2.11}
\begin{equation}
\chi^{(n)} \sim u_n + v_n  p^{d+1-n} \quad,
\label{eq:2.11a}
\end{equation}
where $u_n$ and $v_n$ are finite numbers and we have cut off the infrared
divergence by means of a cut-off momentum {\bf p}. Rotational symmetry
in spin space requires $u_{2n+1}=0$. Again, the nonanalytic
behavior is confined to zero temperature, at finite temperature
$\chi^{(n)}$ is given by
\begin{equation}
\chi^{(n)} \sim u_n + v_n (T + p)^{d+1-n} \quad,
\label{eq:2.11b}
\end{equation}
\end{mathletters}%
which is finite for $p \rightarrow 0$. 
Equations\ (\ref{eq:2.10a}) and\ (\ref{eq:2.11a}) imply that our
LGW functional, Eq.\ (\ref{eq:2.7a}), contains a nonanalyticity which
has the form of a power series in $M/p$. In order to specify the
LGW functional, we still need a physical interpretation of the
infrared cutoff momentum ${\bf p}$. This will be given in the next
section.

To conclude this section we show that the static correlation functions
discussed above provide more important nonanalyticities than higher order
terms in a frequency expansion. To see this, 
we anticipate a result from the next 
section. We will discuss a fixed point of the renormalization group
transformation where the dynamical exponent is $z=d$, i.e. $\Omega_n$
scales like $q^d$. Now let us look at an expansion of $\chi^{(n)}$
in powers of frequency. In our clean system with ballistic
modes the term of $m^{\rm th}$ order in $\Omega_n$ carries an 
additional factor of $q^m$ in 
the denominator, as can be seen in Eq.\ (\ref{eq:2.9}) for the spin 
susceptibility. Thus, if the term of zeroth order in the frequency scales like
$q^{d+1-n}$ near criticality,  then the $m$th order term scales like
$q^{d+1-n+m(d-1)}$. For $d>1$ the static susceptibility therefore has a
stronger divergence than the frequency corrections. However, in $d=1$ the
leading divergence is provided by a frequency dependent term, as will be
discussed in Sec.\ \ref{sec:V} below.

\section{The Critical Behavior at zero temperature}
\label{sec:III}

Here we discuss the quantum critical behavior 
of the Gaussian part of the LGW theory defined in the last section.
We do so both by explicitly solving the Gaussian theory, and by
studying the 
renormalization group properties of the Gaussian fixed point.
We then analyze the non-Gaussian terms in the field theory, and
show that they are irrelevant, in the renormalization group sense, with
respect to the Gaussian fixed point for all dimensions $d>1$.
This implies that the Gaussian theory yields the exact
critical behavior for all of these dimensions, except for logarithmic
corrections to scaling in $d=3$ that are discussed in Appendix \ref{app:B}. 
We then construct the equation
of state near the critical point. This requires a more detailed discussion
of the irrelevant non-Gaussian terms in the field theory, since the
equation of state is determined in part by dangerous irrelevant variables. 

\subsection{The Gaussian fixed point}
\label{subsec:III.A}

According to Eqs.\ (\ref{eqs:2.7}), (\ref{eq:2.9}), and (\ref{eq:2.10a}), 
the Gaussian part of the LGW functional $\Phi[{\bf M}]$ is,
\begin{mathletters}
\label{eqs:3.1}
\begin{eqnarray}
\Phi_2[{\bf M}] &&= {1\over 2} \sum_{\bf q}\sum_{\Omega_n}
     {\bf M}({\bf q},\Omega_n)\,\left[t_0 + a_{d-1}\,q^{d-1} \right.
\nonumber\\
      &&\left. +\ a_2\,q^2 + a_{\Omega}\,\vert\Omega_n\vert/q\right]\,
     \cdot {\bf M}(-{\bf q},-\Omega_n)\ ,
\label{eq:3.1a}
\end{eqnarray}
where 
\begin{equation}
t_0 = 1 - \Gamma_t\,\chi_s({\bf q}\rightarrow 0,\Omega_n=0)\quad,
\label{eq:3.1b}
\end{equation}
\end{mathletters}%
is the bare distance from the critical point, and $a_{d-1}$, $a_2$, and
$a_{\Omega}$ are positive constants.

We first analyze the critical behavior implied by Eqs.\ (\ref{eqs:3.1}).
Later we will show that for $d>1$ fluctuations are irrelevant, and the
critical behavior found this way is exact for these dimensions. Four
critical exponents can be directly read off
Eq.\ (\ref{eq:3.1a}). These are
the correlation length exponent $\nu$, defined by
$\xi\sim t^{-\nu}$, with $\xi$ the correlation length and
$t$ the (renormalized) dimensionless distance from the
critical point; the exponent $\eta$ that determines the wavenumber
dependence of the magnetic (i.e., order parameter) susceptibility at
criticality,
$\langle M_a({\bf q,}0)\,M_a({- \bf q},0)\rangle \sim q^{-2+\eta}$; the
dynamical scaling exponent $z$ that characterizes critical slowing down by
relating the divergence of the relaxation time, $\tau_r$, to that of the
correlation length, $\tau_r \sim \xi^z$; and the exponent $\gamma$ that
describes the $t$-dependence of the static magnetic susceptibility
$\chi_m = \langle M_a(0,0)\,M_a(0,0)\rangle \sim t^{-\gamma}$.
An inspection of Eq.\ (\ref{eq:3.1a}) yields,
\begin{mathletters}
\label{eqs:3.2}
\begin{equation}
\nu=\cases{1/(d-1)& for $1<d<3$\cr%
               1/2& for $d>3$\cr}%
     \quad,
\label{eq:3.2a}
\end{equation}
\begin{equation}
\eta = \cases{3-d& for $1<d<3$\cr%
                0&   for $d>3$\cr}%
        \quad,
\label{eq:3.2b}
\end{equation}
\begin{equation}
z = \cases{d& for $1<d<3$\cr%
           3& for $d>3$\cr}%
        \quad,
\label{eq:3.2c}
\end{equation}
\begin{equation}
\gamma = 1 \quad \mbox{ for all } d>1 \quad.
\label{eq:3.2d}
\end{equation}
\end{mathletters}%
In $d=3$ there are logarithmic corrections to scaling, see 
Appendix\ \ref{app:B}.

For later reference, we also discuss the critical behavior given by
Eqs.\ (\ref{eqs:3.1}) and (\ref{eqs:3.2}) from a renormalization group
point of view. Let $b$ be the renormalization group length rescaling
factor. Under renormalization, all quantities change according to
$A\rightarrow A(b)= b^{[A]}\,A$, with $[A]$ the scale dimension of $A$. The
scale dimension of the order parameter is,
\begin{mathletters}
\label{eqs:3.3}
\begin{equation}
[{\bf M}({\bf q},\Omega_n)] = -1+\eta/2 \quad,
\label{eq:3.3a}
\end{equation}
or, equivalently,
\begin{equation}
[{\bf M}({\bf x},\tau)] = (d+1)/2 \quad.
\label{eq:3.3b}
\end{equation}
\end{mathletters}%
At the critical fixed point, $a_{\Omega}$ and either $a_{d-1}$ (for
$1<d<3$), or $a_2$ (for $d>3$) are not renormalized, i.e. their scale
dimensions are zero. Using this, and $[q]=1$, $[\Omega_n]=z$
immediately yields $\eta$ and $z$ as given by 
Eqs.\ (\ref{eq:3.2b},\ \ref{eq:3.2c}). 
Equation\ (\ref{eq:3.2a}) follows from the relevance of $t_0$, or its
renormalized counterpart, $t$,
at the critical fixed point. That is, the scale dimension of $t$ is
positive and given by $1/\nu\equiv [t] = 2-\eta$.

\subsection{The non-Gaussian terms}
\label{subsec:III.B}

We now show that all of the non-Gaussian terms in the field theory are
renormalization group irrelevant with respect to the Gaussian fixed point
discussed in the last subsection. To do this we determine the scale dimensions
of the coefficients $u_n$ and $v_n$ of the higher-order terms in the 
LGW functional which had been defined in Eq.\ (\ref{eq:2.11a}). Since
the latter have been defined in Fourier space, we first take the Fourier
transform of the $n^{\rm th}$ summand in Eq.\ (\ref{eq:2.7a}). This yields
\begin{eqnarray}
\left[\chi^{(n)}\right] = -{1\over 2}(n-2)[T/V] - 
                                   n\left[M({\bf q},\Omega_n)\right]
\nonumber\\
                        = -{1\over 2}(n-2)(d+z) + n(1-\eta/2)\quad,
\label{eq:3.4}
\end{eqnarray}
where we have used Eq.\ (\ref{eq:3.3a}). Next we need to assign a scale
dimension to the cutoff momentum $p$ in Eq.\ (\ref{eq:2.11a}). The most
obvious guess from a scaling point of view is to identify $p$ with the
inverse correlation length $\xi^{-1}$, which makes $[p]=1$. We will
ascertain in Sec.\ \ref{subsec:III.C} below that this is indeed the
correct choice. It then follows from Eq.\ (\ref{eq:2.11a}) that of
the two parameters $u_n$ and $v_n$, $v_n$ has the larger scale dimension,
and hence is more relevant,
for $d<n-1$, while $u_n$ is more relevant for $d>n-1$. For even $n$ 
(remember that $u_{2n+1} = 0$) and $d>n-1$ the most relevant parameter
thus has a scale dimension
\begin{mathletters}
\label{eqs:3.5}
\begin{equation}
\left[u_{2n}\right] = -(d+1)(n-1) + 2\quad,
\label{eq:3.5a}
\end{equation}
which is always negative. For odd $n$, and for $d<n-1$ for even $n$,
we need to consider,
\begin{equation}
[v_n] = \cases{-(n-2)(d-1)/2& for\ $1<d<3$\cr%
                  2-n(d-1)/2& for\ $d<3$\cr}%
        \quad.
\label{eq:3.5b}
\end{equation}
\end{mathletters}% 
We see that all of the $v_n$ are irrelevant for $d>1$, while in $d=1$
they all become marginal.
We conclude that for $d>1$ the critical Gaussian fixed point is stable,
and so the exponents given in Eqs.\ (\ref{eqs:3.2}) are exact. In $d=1$ an
infinite number of operators seems to become marginal, so naively one
would conclude that the upper critical dimension is given by $d_c^+ =1$.
However, one has to keep in mind that the functional form of correlation
functions in $d=1$ can be qualitatively different from that in $d>1$, so
our power counting may be valid only for $d>1$, and the scale dimension of
some operators may change discontinuously from irrelevant in $d>1$ to
relevant in $d=1$. This actually happens, as will be discussed 
in Sec.\ \ref{sec:V} below.

\subsection{The equation of state}
\label{subsec:III.C}

We now determine the equation of state, and calculate the critical
behavior of the magnetization and the magnetic susceptibility. 
Since we have shown in the last subsection that fluctuations are
irrelevant for the critical behavior for all $d>1$, we can determine
the equation of state by simply calculating the saddle point
contribution to the free energy. However,
in order to do so,
we have to include the higher order terms in the LGW functional.
Although they are irrelevant operators for $d>1$, they are potentially
dangerously irrelevant\cite{MEF} with respect to the magnetization $m$.
We will see below that all of the $v_n$ are indeed dangerouly irrelevant, 
since $m$ is a singular function of the $v_n$ for $v_n \rightarrow 0$. 

We determine the mean field
equation of state by calculating the saddle point 
contribution to the free energy. To do so, we replace the order parameter
field ${\bf M}(x)$ in the LGW functional, Eq.\ (\ref{eq:2.7a}), by the
average magnetization $m$ and determine the stationary point of $\Phi$
with respect to $m$. This yields the equation of state in the form
\begin{equation}
t\, m + u_4\, m^3 + m^d \sum_{n=3}^{\infty} v_n\,(m/p)^{n-1-d} = H \quad,
\label{eq:3.6}
\end{equation}
where an external magnetic field $H$ was added. We have only kept the 
leading terms in each order of $m$, and we have suppressed all numerical 
prefactors since they are unimportant for our purposes. The equation of
states now has the form of a power series in $m/p$. This implies that
the cut-off momentum $p$ effectively scales like $m$.
Thus, all higher order terms effectively have the same power of $m$, 
viz. $m^d$, and the equation of states now reads
\begin{equation}
t\,m + u_4\,m^3 + v\, m^d = H\quad,
\label{eq:3.7}
\end{equation}
with $u_4$ from Eq.\ (\ref{eq:2.11a}) and $v$ another finite coefficient.
From Eq.\ (\ref{eq:3.7})
we immediately obtain the exponents $\beta$ and $\delta$, defined by
$m(t,H=0)\sim t^{\beta}$, $m(t=0,H)\sim H^{1/\delta}$, as
\begin{mathletters}
\label{eqs:3.8}
\begin{equation}
\beta = \cases{1/(d-1)& for $1<d<3$\cr%
                   1/2& for $d>3$\cr}%
     \quad,
\label{eq:3.8a}
\end{equation}
\begin{equation}
\delta = \cases{d& for $1<d<3$\cr%
                3& for $d>3$\cr}%
     \quad.
\label{eq:3.8b}
\end{equation}
\end{mathletters}%
In $d=3$ logarithmic corrections to scaling occur, see Appendix \ref{app:B}.

We are now in a position to determine the exact scale dimension of the 
cut-off momentum $p$ and to verify the idetification of $p$ with
$\xi^{-1}$ made in the last subsection.
As we have seen after 
Eq.\ (\ref{eq:3.6}), the cutoff $p$ scales like $m$, so that
$[p]=[m]_{\rm eff}$, where $[m]_{\rm eff}=\beta/\nu$ is the {\it effective}
scale dimension of $m$, i.e. the scale dimension after the effects of
the dangerous irrelevant variables have been taken into account.
From Eqs.\ (\ref{eq:3.8a}), (\ref{eq:3.2a}) we see that
$\beta = \nu$, hence $[m]_{\rm eff} = 1$.
Consequentely, $[p]=1$ which justifies the identification 
$p$ with $\xi^{-1}$ for scaling purposes made in Sec.\ \ref{subsec:III.B}.
These results complete the proof of the statement that the system is above 
its upper critical dimension for $d>1$.

\section{Behavior at finite temperatures, and the quantum to classical 
                                                               crossover}
\label{sec:IV}

In the first part of this section we discuss various sources of 
temperature dependence in our field theory and the corresponding scaling
behavior of the temperature. In the following two subsections we apply
the results to a calculation of the equation of state at finite temperatures
and of the specific heat, and we discuss the crossover from the quantum 
Gaussian fixed point to the classical Gaussian fixed point. In the last
subsection we discuss the analogous temperature dependence in systems
with quenched disorder.

\subsection{Scaling behavior of the temperature}
\label{subsec:IV.A}

The temperature dependence of the LGW functional, Eq.\ (\ref{eq:2.7a}), 
is due to a number of entirely different effects. First, the spin-density 
correlation functions $\chi^{(n)}$ that determine the coeffients of the LGW
functional are temperature dependent as given in Eqs.\ (\ref{eq:2.10b}) and\
(\ref{eq:2.11b}). These are correlation functions for the reference
ensemble that is far from any critical point. Therefore, the frequency
or temperature in $\chi^{(n)}$ scales like the momentum, and consequentely 
we have a temperature scale whose scale dimension is,
\begin{mathletters}
\label{eqs:4.1}
\begin{equation}
[T]_{\rm ball} = 1\quad. 
\label{eq:4.1a}
\end{equation}
This we will refer to as the ballistic temperature scale.

A second source of temperature dependence is the usual dynamical scaling.
This
originates from the fact that the time integration in the LGW functional,
Eq.\ (\ref{eq:2.7a}), extends over the finite interval $[0,1/T]$. 
Dynamical scaling is hence equivalent
to finite-size scaling in the direction of imaginary time. The scaling 
behavior of the temperature due to dynamical scaling is described by the 
dynamical exponent $z$ as given in Eq.\ (\ref{eq:3.2c}). We therefore
have a second temperature scale whose scale dimension is 
\begin{equation}
[T]_{\rm crit} = z\quad,
\label{eq:4.1b}
\end{equation}
which below we will refer to as the critical temperature scale.

Finally, the distance from the critical point, $t$, is temperature
dependent. For the bare $t_0$, this is simply the usual $T^2$ dependence
that is familiar from Fermi liquid theory. The scale dimension of this
Fermi liquid temperature scale is thus 
\begin{equation}
[T]_{\rm FL} = 1/2\nu\quad.
\label{eq:4.1c}
\end{equation}
However, upon renormalization $t$ acquires a more intricate temperature 
dependence. Since the loop or fluctuation
corrections involve integrals over critical propagators, the latter depends on
the critical temperature scale. However, it also depends on the scale
dimensions of the vertices $u_n$ and $v_n$, Eq.\ (\ref{eq:2.11a}), that
are dangerous irrelevant variables with respect to the $T$-dependence of
the magnetization and the magnetic susceptibility.
Millis\cite{Millis} and Sachdev\cite{SachdevT} have shown that the resulting
temperature scale has a scale dimension 
\begin{equation}
[T]_{\rm fluct} = z/(1-\nu\theta)\quad,
\label{eq:4.1d}
\end{equation}
\end{mathletters}%
where $\theta$ is the scale dimension of the appropriate dangerous irrelevant
operator, $u_4$ in our case. This we call the fluctuation temperature scale.

Of all the temperature scales that are present in a given quantity, the one
with the largest scale dimension will be the dominant one. This means, for
instance, that the Fermi liquid scale will be subdominant compared to the
ballistic one for all $1<d<3$; for $d>3$ the two scales are indistinguishable.
Since $z>1$ for all $d>1$ it also means that the critical temperature scale
will in general be dominant over both the ballistic scale and the 
fluctuation scale. There are, however,
two possible mechanisms that can invalidate this conclusion.
The first possibility is
that some quantities do not depend on the critical temperature
scale. For those the leading temperature dependence will be given by
either the ballistic temperature scale, or by the fluctuation scale,
depending on the values of $[T]_{\rm ball}$ and $[T]_{\rm fluct}$. 
As we will see, this possibility is
realized for the magnetization and the magnetic susceptibility.
The second possibility is that a subdominant
temperature scale is dangerously irrelevant. This possibility cannot be
ruled out by general scaling considerations.

\subsection{Magnetization, and magnetic susceptibility}
\label{subsec:IV.B}

Since we work above an upper critical dimensionality, the 
magnetization $m$ and the magnetic susceptibility $\chi_m$ are
determined by the $q=\Omega_n=0$ Fourier component of the order
parameter field, and do not
depend on the finite-frequency behavior of the critical modes.
Therefore, dynamical scaling does {\em not} enter the temperature
dependence of $m$ and $\chi_m$ directly. In other words, their behavior
is completely 
determined by the equation of state, Eq.\ (\ref{eq:3.6}), whose 
coefficients acquire finite temperature corrections according to 
Eqs.\ (\ref{eq:2.10b}) and\ (\ref{eq:2.11b}). The $t\,m$ term in
Eq.\ (\ref{eq:3.6}) has a correction of the form
$m\, (p+T)^{d-1} \sim m\, (m+T)^{d-1}$. Similarly, the higher order
terms $v_n\, p^{d+1-n}\, m^{n-1}$ have corrections 
proportional to $(p+T)^{d+1-n}\, m^{n-1} \sim (m+T)^{d+1-n}\, m^{n-1}$.
This is the ballistic temperature scale discussed in Sec.\ \ref{subsec:IV.A}
above. In addition, there is the fluctuation temperature scale that
results from the temperature dependence of $t$. By using Eq.\ (\ref{eq:3.5a})
in Eq.\ (\ref{eq:4.1d}) we see that $[T]_{\rm ball} > [T]_{\rm fluct}$, 
and thus is
the dominant temperature scale, for $1<d<2$, and for $d>5$, but that for
$2<d<5$, $[T]_{\rm fluct}$ is the relevant scale.
All of these corrections can be summarized in the following 
scaling law for the magnetization $m$,
\begin{mathletters}
\label{eqs:4.2}
\begin{equation}
m(t,T,H) = b^{-\beta/\nu} m(tb^{1/\nu}, Tb^{\phi/\nu}, Hb^{\delta\beta/\nu})
                                                                    \quad,
\label{eq:4.2a}
\end{equation}
where $b$ is an arbitrary scale factor. The crossover exponent
\begin{equation}
\phi = \cases{\nu = 1/(d-1)\quad, &for $1<d<2$\quad,\cr
              d/2(d-1)\quad,      &for $2<d<3$\quad,\cr
              3/(d+1)\quad,       &for $3<d<5$\quad,\cr
              \nu = 1/2\quad,     &for $d>5$\quad,\cr}
\label{eq:4.2b}
\end{equation}
\end{mathletters}%
describes the crossover from the quantum critical region to a regime whose 
behavior is dominated by the classical Gaussian fixed 
point.\cite{Millis,crossoverfootnote} Note that, for the reasons
explained above, the crossover exponent is {\em not} given by $z\nu$ which
one would expect from dynamical scaling. Also note the complicated
behavior of the crossover exponent $\phi$ as a function of the dimensionality,
which is brought about by the competition between the ballistic and the
fluctuation temperature scales. The result $\phi=\nu$ that was reported
in Refs.\ \onlinecite{Letter} and \onlinecite{fm_sb} was correct only for
$1<d<2$ and $d>5$. By differentiating 
Eq.\ (\ref{eq:4.2a}) with respect to $H$, we obtain the analogous 
homogeneity law for the magnetic susceptibility $\chi_m$,
\begin{equation}
\chi_m(t,T,H) = b^{\gamma/\nu} \chi_m(tb^{1/\nu}, Tb^{\phi/\nu}, 
                           Hb^{\delta\beta/\nu})\quad.
\label{eq:4.3}
\end{equation}

\subsection{The specific heat}
\label{subsec:IV.C}

The scaling behavior of the specific heat is determined by the sum of
the mean-field and the Gaussian fluctuation contribution to the
free energy density $f$. The mean-field contribution follows immediately from
Eq.\ (\ref{eq:3.7}). The Gaussian fluctuation contribution, $f_G$, 
which gives the leading nonanalyticity for the specific heat at the 
critical point, can be calculated in complete analogy to the case of
classical $\phi^4$-theory in $d>4$.\cite{Ma} Neglecting an uninteresting
constant contribution to $f_G$, we obtain,
\begin{eqnarray}
f_G = {T\over 2V} \sum_{{\bf q},\omega_n} \biggl[2 \ln\left(
       H/m + a_{d-1}\,q^{d-1}\right.
\nonumber\\
       \left. + a_2\,q^2 + a_{\Omega}\,\vert\Omega_n\vert
       /q\right)
\nonumber\\
      + \ln\left(x_d\,H/m - (x_d-1)\,t + a_{d-1}\,q^{d-1}\right.
\nonumber\\
      \left. + a_2\,q^2 + a_{\Omega}\,
           \vert\Omega_n\vert/q\right)\biggr]\quad.
\label{eq:4.4}
\end{eqnarray}
Here $x_d = d$ for $1<d<3$, and $x_d=3$ for $d>3$. 
The specific heat coefficient $\gamma_V$ is conventionally defined by,
\begin{equation}
\gamma_V = c_V/T = -\partial^2 f/\partial T^2\quad.
\label{eq:4.5}
\end{equation}
Again we are interested only in scaling properties and not in exact
coefficients. Keeping this in mind, an adequate representation of
Eqs.\ (\ref{eq:4.4}) and (\ref{eq:4.5}) is given by the integral,
\begin{equation}
\gamma_V = \int_0^{\Lambda} dq\ {q^{d-1}\over T+q^d+q^3+Hq/m}\quad,
\label{eq:4.6}
\end{equation}
with $\Lambda$ an ultraviolet cutoff.

Let us point out two interesting features of this result. 
First, for all dimensionalities in the range $1<d\leq 3$,
$\gamma_V$ is logarithmically singular for $T,H\rightarrow 0$. This can
be seen most easily from Eq.\ (\ref{eq:4.6}), and it is also true for
the exact result, Eqs.\ (\ref{eq:4.5}), (\ref{eq:4.4}).
Such a $d$-independent logarithmic singularity is somewhat unusual.
Wegner has discussed how logarithmic corrections to scaling arise
if a set of scale dimensions fulfills some resonance condition.\cite{Wegner}
In the present case the relevant relation is
that the scale dimension of the free energy, $d+z=2z$ for $1<d<3$,
is a multiple of the scale dimension of $T$, which is $z=d$ in this region.
The fact that the logarithm appears in a {\it range} of
dimensions, rather than only for a special value of $d$, is due to the
dynamical exponent being exactly $d$ in that range.
Second, as discussed in Sec.\ \ref{subsec:IV.A}, two different 
temperature scales appear in Eq.\ (\ref{eq:4.6}).
The first two terms in the denominator indicate that $T\sim\xi^{-d}$, as
one would expect from dynamical scaling. However, the last term in
Eq.\ (\ref{eq:4.6}) contains the magnetization, which in turn depends
on $[T]_{\rm ball}$ and $[T]_{\rm fluct}$. These two features
imply that the scaling equation for $\gamma_V$ should be written,
\begin{eqnarray}
\gamma_V(t,T,H) = \Theta(3-d)\,\ln b \qquad\qquad\qquad\qquad
\nonumber\\
            + F_{\gamma}(t\,b^{1/\nu},T\,b^z,
                  H\,b^{\delta\beta/\nu},T\,b^{\phi/\nu})\quad.
\label{eq:4.7}
\end{eqnarray}
Note that the scale dimension of $\gamma_V$, ignoring the logarithm,
is zero in all dimensions.
Since $z>\phi/\nu$ for all $d>1$, one can formally ignore the fourth 
entry in the scaling
function since it is subleading compared to the second entry and
its effects can be considered as `corrections
to scaling'. We emphasize that in contrast to the magnetization and the magnetic
susceptibility, the specific heat {\em does} depend on the critical modes, and
hence contains the critical temperature scale. As mentioned in 
Sec.\ \ref{subsec:IV.A}, the latter is dominant when it is present, 
and $\gamma_V$ provides an example for that.

\subsection{The disordered case revisited}
\label{subsec:IV.D}

Let us finally reconsider the case of systems with quenched 
disorder.\cite{fm} We do this partly to point out the remarkable analogy
between the clean and dirty cases, and partly to correct a mistake in
the results of Ref.\ \onlinecite{fm} for dimensions $d>3$.

In the presence of disorder, the temperature or frequency far from 
criticality scales like the wavenumber squared. The ballistic temperature
scale of Eq.\ (\ref{eq:4.1a}) therefore gets replaced by a diffusive one,
\begin{equation}
[T]_{\rm diff} = 2\quad.
\label{eq:4.8}
\end{equation}
The scale dimensions for the other temperature scales, Eqs.\ (\ref{eq:4.1b})
- (\ref{eq:4.1d}) remain valid, but the values of the exponents $z$, $\nu$, and
$\theta$ change. In the action, the disorder leads to
terms in addition to, and structurally different from, those in 
Eq.\ (\ref{eq:2.7a}).\cite{fm}
In particular, at order ${\bf M}^4$ a second term appears whose coupling
constant was denoted by $v_4$ in Ref. \onlinecite{fm}. This operator,
whose scale dimension is $[v_4] = - \vert d-4\vert$, is the least irrelevant
operator that is dangerous with respect to $t$, and one therefore has
$\theta = - \vert d-4\vert$ in the fluctuation temperature scale,
Eq.\ (\ref{eq:4.1d}). With $z$ and $\nu$ as determined in 
Ref.\ \onlinecite{fm}, this leads to a crossover exponent
\begin{equation}
\phi = \cases{2\nu = 2/(d-2)\quad, &for\quad $2<d<\sqrt{5}+1$\quad,\cr
              d/2\quad,            &for\quad $\sqrt{5}+1<d<4$\quad,\cr
              4/(d-2)\quad,        &for\quad $4<d<6$\quad,\cr
              2\nu = 1\quad,       &for\quad $d>6$\quad,\cr}
\label{eq:4.9}
\end{equation}
\par\noindent
The result $\phi = 2\nu$ of Refs.\ \onlinecite{fm,fm_sb} was thus not
correct for the (unphysical) dimensionality range $\sqrt{5}+1<d<6$. The scaling
behavior of the magnetization and the magnetic susceptibility is given
by Eqs.\ (\ref{eq:4.2a}) and (\ref{eq:4.3}), respectively, with $\phi$
from Eq.\ (\ref{eq:4.9}), and all other exponents as given in
Ref.\ \onlinecite{fm}.

\section{Hertz's model revisited}
\label{sec:V}

In this section we reexamine Hertz's original model of the 
ferromagnetic quantum phase transition.\cite{Hertz} We show that at
tree level, the LGW theory for this model breaks down for related, but
somewhat different reasons than in the realistic model above. Moreover,
starting at one-loop order the renormalization group generates terms
that are not in the bare action. As a result, the critical behavior
of this model in $d>1$ is {\em not} mean-field like, but rather the same as
that of the more realistic model we have studied so far.

Hertz's model differs in two respects from the more realistic one given
by Eqs.\ (\ref{eq:2.3}) through (\ref{eqs:2.5}). 
First, the interaction part of the action contains only the particle-hole
spin-triplet channel that is decoupled in the derivation
of the LGW functional. Consequentely, the reference ensemble $S_0$ 
consists of noninteracting electrons. Second, the coefficients
$\chi^{(n)}$ of the LGW functional are taken to be the correlation
functions of a $3$-$d$ Fermi gas, irrespective of the dimensionality
of the space the magnetization is confined to. For this model the 
spin susceptibility $\chi_s$ of the reference system is simply the 
Lindhardt function, so Eq.\ (\ref{eq:2.9}) gets replaced by
\begin{equation}
\chi_s({\bf q}, \Omega_n) = c_0 - c_2\, q^2 - \vert \Omega_n \vert /q 
\quad.
\label{eq:5.1}
\end{equation}
In comparison to the analogous expression for an interacting Fermi liquid, 
Eqs.\ (\ref{eq:2.9}) and\ (\ref{eq:2.10a}), the term proportional to
$q^{d-1}$ is missing, and $\chi_s({\bf q}, \Omega_n=0)$ is now analytic 
at ${\bf q}=0$. The resulting Gaussian part of the bare LGW functional 
has the form
\begin{eqnarray}
\Phi_2[{\bf M}] = {1\over 2} \sum_{\bf q, \Omega_n}
     {\bf M}({\bf q},\Omega_n)\,\left[t_0  
        +\ a_2\,q^2 + a_{\Omega}\,\vert\Omega_n\vert/q\right]
\nonumber\\
     \times {\bf M}(-{\bf q},-\Omega_n) \quad.
\label{eq:5.2}
\end{eqnarray}
This action allows for a Gaussian fixed point with mean-field static 
exponents and a dynamical exponent $z=3$.\cite{Hertz}

Let us now investigate the stability of this Gaussian fixed point. At tree
level, we can do this by calculating the scale dimensions of the 
coefficients of the higher
order terms. Hertz considered the higher order correlation functions 
$\chi^{(n)}$ only in the limit $\Omega_n =0$, $q \rightarrow 0$ where
they are finite numbers. The usual power counting arguments in
analogy to Sec.\ \ref{subsec:III.B} show that all of those terms
are irrelevant for $d>1$. The quartic one becomes marginal in $d=1$
and relevant for $d<1$. This changes, however, if one considers
the first order term in an 
expansion of the $\chi^{(n)}$ for non-interacting electrons in 
powers of frequency. The reason is that, at nonzero external frequency,
frequency mixing effects occur that are similar to those brought about
by an electron-electron interaction (which causes them even at zero 
external frequency).
For power counting purposes, i.e. to determine the scale dimensions of the 
coefficients, it is not necessary
to calculate the $\chi^{(n)}$ completely as functions of $n-1$ frequencies
and wavevectors. Guided by more complete calculations for the
cases $n=3,4,5$, we have concluded 
that for power counting purposes it
suffices to consider one independent frequency, $\Omega$, and two
directionally independent wavevectors of equal length,
$\vert{\bf q}_1\vert = \vert{\bf q}_2\vert = \vert{\bf q}\vert$, that
form an angle $\theta$. With these simplifications the leading term of
the general coefficient
is easily calculated. We find that for odd $n$ the linear-in-frequency
term in the action can symbolically be written
\begin{mathletters}
\label{eqs:5.3}
\begin{equation}
w_n \int dx\ M^n(x)\ \Omega/q^{n-1}\quad,
\label{eq:5.3a}
\end{equation}
with $w_n$ some coupling constant, while the corresponding terms for
even $n$ are less relevant. The same power counting arguments that we
used in Sec.\ \ref{subsec:III.B} show that the scale dimension
of $w_n$ is $[w_n] = -(d-1)(n-2)/2$, which is negative for all
$d>1$. The coupling constants $w_n$
are thus irrelevant for all physical dimensions
$d\geq 2$, and seem to become marginal in $d=1$. However,
Eq.\ (\ref{eq:5.3a}) holds only if the two independent wavevectors
are neither parallel nor antiparallel, i.e. it holds only if
$d >1$. For $\theta = 0, \pi$ one
finds a stronger singularity for the terms with even $n$,
\begin{equation}
w_n \int dx\ M^n(x)\ \Omega/ q^{2n-3}\quad,\quad,
\label{eq:5.3b}
\end{equation}
\end{mathletters}%
while the terms with odd $n$ are less relevant. In dimensions
$d >1$ the parallel wavevectors form a set of measure zero and this
stronger singularity is of no consequence. In $d=1$, however, 
Eq.\ (\ref{eq:5.3b})
represents the generic behavior of the terms of $O(M^n)$. Power counting
yields the scale dimensions to be
$[w_n] = n-2$ in $d=1$, so all of these coefficients are relevant operators.
This is sufficient to conclude that the upper critical dimension is {\it not}
one, but rather that the $1$-$d$ sytem is {\it below} its upper critical
dimension, and will show nontrivial critical behavior. This provides a
technical explanation for Sachdev's observation\cite{Sachdev} that Hertz's 
results in $d<1$ cannot be correct.

\begin{figure}
\epsfxsize=8.5cm
\epsfysize=5.6cm
\epsffile{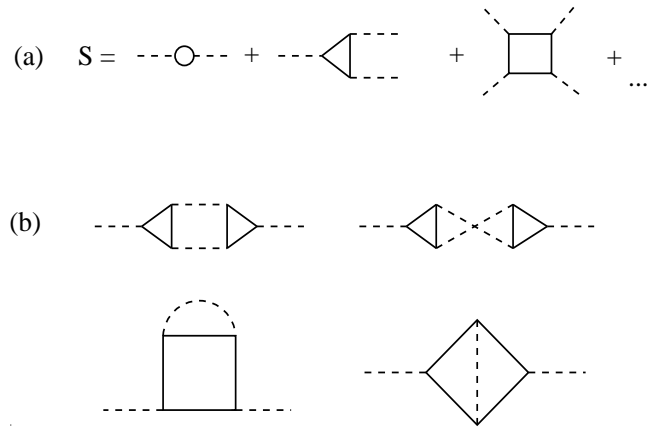}
\vskip 0.5cm
\caption{(a) Symbolic representation of the action. The dashed lines represent
 the external legs, and the polygons represent the vertices. 
 (b) Renormalization of the two-point vertex to one-loop order.}
\label{fig:V.1}
\end{figure}

Moreover, the renormalization of the model beyond the tree level qualitatively
changes the form of the Gaussian action, Eq.\ (\ref{eq:5.2}). Consider, for
instance, the one-loop renormalization of $\Phi_2$ by the terms of order
${\bf M}^3$ and ${\bf M}^4$. The corresponding diagrams are shown in
Fig.\ \ref{fig:V.1}. It is easy so see that these diagrams are equivalent
to those that determine the spin susceptibility of interacting electrons,
and have been calculated in Ref.\ \onlinecite{chi_s}.\cite{diagramsfootnote}
Renormalization therefore leads to a term proportional to $q^{d-1}$ in the
Gaussian action, and hence to a $\Phi_2$ as given by Eq.\ (\ref{eq:3.1a}).
We conclude that the critical behavior of Hertz's model for $d>1$ is {\em not}
mean-field like, but rather the same as that of the more realistic model
dicussed in Secs.\ \ref{sec:II} - \ref{sec:IV}. 

\section{Discussion}
\label{sec:VI}

In this paper we have shown that clean itinerant quantum
ferromagnets at zero temperature do not show, as was previously thought, 
uninteresting mean-field critical behavior for all dimensionalities $d>1$.
Rather, there are two upper critical dimensions. The first one is $d_c^+ = 1$,
above which the critical behavior is described by a Gaussian field theory
but is not mean-field like, and the second one is $d_c^{++} = 3$, above which
one does find mean-field critical behavior. The reason for this unusual
behavior is soft modes that lead to an effective long-range interaction
between the order parameter fluctuations. As is the case for classical models 
with long-range interactions,\cite{FisherMaNickel} this leads to nontrivial
critical behavior that still can be determined exactly. $d_c^{++}$ is the
marginal dimension where the soft mode induced long-range interaction
coincides with the usual $r^{-(d+2)}$ behavior. In this final section
we discuss a few aspects of these results that have not been covered yet.

First of all, both our approach and our results are remarkably similar to
a recent treatment of {\em disordered} itinerant quantum ferromagnets.\cite{fm}
In these papers it was found that the disorder induced diffusive excitations
in a Fermi system with quenched disorder lead to similar, but stronger effects,
with a long-ranged interaction between order parameter fluctuations that
falls off like $r^{-(2d-2)}$, and {\em three} `upper critical 
dimensionalities', viz. $2$, $6$, and $4$. The first two are analogous to
$d_c^+$ and $d_c^{++}$ above. In addition, the critical exponents $\nu$,
$\eta$, and $z$ take on their mean-field values for $d>4$, while $\beta$ and
$\delta$ remain anomalous between $d=4$ and $d=6$. In the present case, a
different structure of the dangerous irrelevant variables makes the upper
two special dimensions coincide. The same difference in the dangerous
irrelevant variables leads to a difference in the temperature dependence
of the equation of state as a function of dimensionality: In the disordered
case, the diffusive temperature scale with $[T]_{\rm diff} = 2$, which is
analogous to the ballistic temperature scale $[T]_{\rm ball} = 1$ in
the clean case, is dominant in the physical dimension $d=3$. In the clean 
case, in contrast, the ballistic scale is subdominant compared to the
fluctuation temperature scale for $2<d<5$.

The present paper shows that the basic concepts of Refs.\ \onlinecite{fm}
are not restricted to disordered systems. Indeed, an attempt to construct
an effective field theory entirely in terms of the order parameter field
for any phase transition will break down (in the sense that it is impossible
to construct a {\em local} effective theory) if there are
soft or slow modes other than the order parameter fluctuations that couple 
to the order parameter. In the present case, the spin-triplet particle-hole 
excitations that always exist in an interacting Fermi system, and that are
distinct from the order parameter mode in that they are soft even in the
paramagnetic phase, are such additional soft modes.
They lead to the nonanalytic behavior of spin density correlation functions
that is displayed in Eqs.\ (\ref{eqs:2.10}) and (\ref{eqs:2.11}), and
hence to the effective long-ranged interaction between the order parameter
modes. In general, the appearance of such additional soft modes would call
for the derivation of a different effective theory that does not integrate
out as many degrees of freedom, and that keeps all of the soft modes on
equal footing. However, as in Ref.\ \onlinecite{fm} we have opted here
to work with the order parameter effective theory after all, since it
turns out that the difficulties introduced by the nonlocal character of
the field theory can be overcome. Nevertheless it would be interesting to
treat the same problem by means of a local theory that keeps more degrees
of freedom explicitly.

It should also be pointed out that our results 
depend crucially on the fact that the
electronic spin density is a conserved quantity. If this feature was
lost, e.g. due to some type of spin flip process, then the spin-triplet
particle-hole excitations would acquire a mass or energy gap, and at
scales larger than this mass the effective interactions between the
order parameter fluctuations would be of short range. The asymptotic
critical phenomena would then be described by a local order parameter
field theory with mean-field critical behavior in all physical dimensions.
At this point one might wonder whether the magnetization in the ordered
phase, and magnetic fluctuations in the disordered one, act effectively
as magnetic impurities, and why this does not lead to an energy gap that
invalidates our conclusions. The answer it that this effect has already
been taken into account. In the ordered phase, the magnetization indeed
acts as a cutoff, as has been discussed in connection with Eq.\ (\ref{eq:3.6}),
and this leads to the nonanalyticity in the equation of state. In the
disordered phase, the cutoff enters only via fluctuations, which are
RG irrelevant with respect to the Gaussian fixed point. The effect
therefore manifests itself only in the corrections to scaling, not in the
leading scaling behavior.

We also mention that all of the qualitative points discussed in
Refs.\ \onlinecite{fm} and \onlinecite{fm_sb} 
that had to do with the fact that one works above
an upper critical dimension, apply here as well. In particular, the
presence of dangerous irrelevant variables means that some of Sachdev's
general results\cite{Sachdev} are not directly applicable to the transition
discussed here. For instance, the Wilson ratio,
$W = (m/H)/(C_V/T)$, diverges at criticality, as it does in the disordered
case, rather than being a universal number as would be the case in the
presence of hyperscaling. However, due to the different structure of the
dangerous irrelevant variables that was already mentioned above in connection
with the multiple upper critical dimensionalities, some details are different
between the disordered and clean cases. For instance, the scaling function
$F_{\gamma}$, Eq.\ (\ref{eq:4.7}), is a function of $T/H$ (if one neglects
the subleading ballistic temperature scale) in agreement with the prediction of
Ref.\ \onlinecite{Sachdev}, while in the presence of disorder this is not
the case.\cite{fm}

We add one more remark concerning Hertz's original model that
was discussed in Sec.\ \ref{sec:V}. Our conclusion that a proper 
renormalization of that model leads to a critical behavior that is the
same as that of the realistic model solves the following paradox that
would otherwise arise: Suppose one divided the interaction term in Hertz's
model into two structurally identical pieces, one carrying some fraction of the
interaction amplitude $\Gamma_t$, and the other the rest. Suppose one then
applied the Hubbard-Stratonovich decoupling only to one of these pieces,
and lumped the other into the reference ensemble. Then the reference
ensemble would contain a nonzero $\Gamma_t$, and according to 
Ref.\ \onlinecite{chi_s} the Gaussian action would contain the nonanalytic
$q^{d-1}$ term that leads to non-mean field critical behavior for all
$d<3$. If one decouples all of the interaction term, on the other hand,
then the reference ensemble has $\Gamma_t=0$, and in the absence of any
other interaction amplitudes there is no way to generate a nonzero
$\Gamma_t$ by renormalization. If Hertz's model had indeed a critical
behavior that is different from that of the realistic model, then the
inevitable conclusion would be that the result depends on how exactly
one performs the decoupling of the interaction term, which would be
physically absurd.

Finally, it should be mentioned that an 
experimental corroboration or refutation of our results is probably
harder for the clean
case discussed in the present paper than for the disordered case
treated in Refs.\ \onlinecite{fm}. There are several reasons for this.
First of all, the zero-temperature behavior can of course not be observed
directly. An experimental study would therefore have to concentrate on the
crossover from the quantum critical behavior to the classical one that will
occur if the classical transition point is at a sufficiently low temperature
for the crossover point to be within the critical region. This requires a
ferromagnet with as low a Curie temperature as possible. In addition, the
parameter $t$ that measures the distance from the critical point is, in
the quantum case, {\em not} the temperature but rather the exchange interaction
or some other microscopic parameter that is hard to control. Both of these
difficulties can be overcome relatively easily in the disordered case, where
a change in the composition of a ferromagnetic alloy changes both the
classical transition temperature and $t$. The quantum critical point is
reached in the low-temperature limit in a sample whose composition is such 
that the Curie temperature just vanishes. 
As was discussed in Ref.\ \onlinecite{fm}, this provides a manageable handle
on $t$ that has no obvious analog in the clean case. Furthermore, the
differences between the quantum critical exponents in $d=3$ and the classical
Heisenberg exponents are more pronounced in the disordered case than in the
clean one. As we have seen, the $3$-$d$ critical behavior in the latter case
is mean-field with logarithmic corrections to scaling. The logarithms would
be hard to observe in any case, and the mean-field exponents are much closer
to $3$-$d$ Heisenberg exponents than those obtained in Ref.\ \onlinecite{fm}.

\acknowledgments

This work was supported by the NSF under grant numbers 
DMR-96-32978 and DMR-95-10185, by the DAAD, by the DFG under grant No.
Vo 659/1-1, and by NATO under grant No. CRG-941250. We would like to thank 
Andy Millis and Subir Sachdev for helpful discussions. DB would like to 
thank Bernhard Kramer for his hospitality at the 
University of Hamburg.

\appendix
\section{Renormalization of the particle-hole spin-triplet interaction vertex}
\label{app:A}

In this appendix we show that a nonvanishing particle-particle channel
interaction, $\Gamma_c\neq 0$, generates a particle-hole spin-triplet
interaction $\Gamma_t$, even if $\Gamma_t = 0$ in the bare system. 

\begin{figure}
\epsfxsize=8cm
\epsfysize=9cm
\epsffile{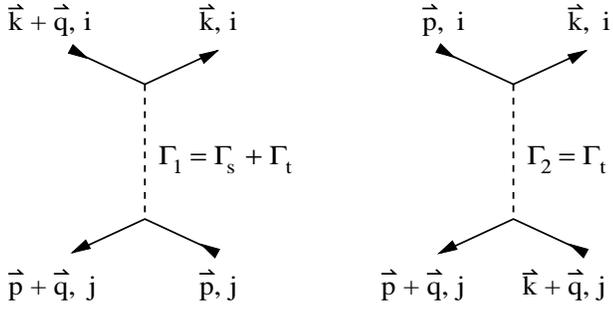}
\vskip 0.5cm
\caption{The three basic interaction vertices. Straight lines with arrows denote
 particles and holes, wavy lines denote the interaction. The notation is
 the same as used in the text.}
\label{fig:A.1}
\end{figure}

\begin{figure}
\epsfxsize=8.25cm
\epsfysize=3.5cm
\epsffile{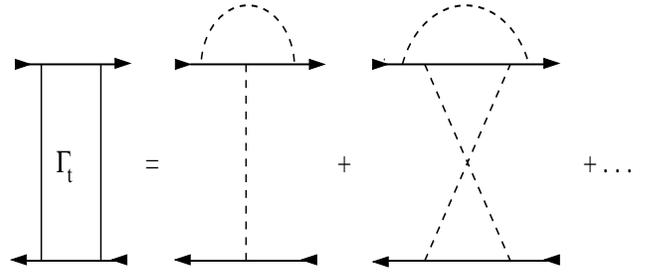}
\vskip 0.5cm
\caption{Construction of a particle-hole spin-triplet vertex out of
 particle-particle vertices.}
\label{fig:A.2}
\end{figure}

Figure \ref{fig:A.1} shows the three basic types of interaction vertices.
Now suppose that $\Gamma_t =0$. Then a vertex of this type can be
constructed by means of the diagrams shown in Fig. \ref{fig:A.2}, where
all of the dashed lines represent Cooper channel interaction amplitudes.

\section{Logarithmic corrections to scaling for $d=3$, and for $1<d<3$}
\label{app:B}

There are three distinct mechanisms that produce logarithmic corrections
to scaling: (1) Marginal operators, (2) Wegner resonance conditions
between a set of scale dimensions, and (3) logarithmic corrections to
the scale dimension of a dangerous irrelevant operator. The first two
mechanism are well known.\cite{Wegner} The third is operative only above
an upper critical dimension, and is therefore of particular interest for
quantum phase transitions.

In the present case, logarithmic corrections to scaling arise due to 
mechanisms (2) and (3). Mechanism (2) produces corrections to the
scaling of the specific heat in all dimensions $1<d\leq 3$, as was discussed
in Sec.\ \ref{subsec:IV.C}. The third mechanism produces corrections
to scaling in $d=3$. According to Eq.\ (\ref{eq:2.11a}), the correlation
function $\chi^{(4)}$ contains a term $v_4\,\ln p$ in $d=3$.
Via Eq.\ (\ref{eq:3.6}) or (\ref{eq:3.7}) this
leads, for instance, to a leading behavior of the spontaneous magnetization,
\begin{equation}
m(t,H=0) \sim {t^{1/2}\over \sqrt{\ln(1/t)}}\quad,
\label{eq:D.1}
\end{equation}
and at the critical point we have
\begin{equation}
m(t=0,H) \sim {H^{1/3} \over \left[\ln(1/H)\right]^{1/3}}\quad.
\label{eq:D.2}
\end{equation}


\begin{references}
\bibitem{LGWfootnote} We use the term `LGW theory' in the narrow sense, in
 which it is usually used in the literature, of
 an effective field theory in terms of the order parameter field only. If
 one defines it as an effective theory for all soft modes, then it is valid
 in the present case, too.
\bibitem{decompositionfootnote} This becomes clearer in Fourier space, where
 the sums over wavenumbers that replace the real space integrals in
 Eq.\ (\ref{eq:2.5a}) extend over small wavenumbers only. See, e.g.,
 Ref.\ \onlinecite{chi_s} for a more detailed exposition of this point.
\bibitem{Landaufootnote} Our interaction constants $\Gamma_s$, $\Gamma_t$, and
 $\Gamma_c$ are the zero angular momentum ($l=0$) 
 components in a multipole expansion
 of the respective interaction potentials, and hence are related to the 
 $l=0$ Landau parameters. The physics we are interested in is due to
 hydrodynamic effects, which are strongest in the $l=0$ or density channel.
 This justifies our neglecting the higher Fermi liquid parameters.
\bibitem{limitsfootnote} This can be seen as follows. Since the magnetization
 is conserved, ordering on a length scale $L$ requires some spin density to be
 transported over that length, which takes a time $t\sim L/v_F$, with $v_F$
 the Fermi velocity. Now suppose the coherence length is $\xi$, and we look
 at the system at a momentum scale $q$ or a length scale 
 $L\sim 1/q < \xi$. Because of the time it takes the system to order over
 that scale, the condition for criticality is $L < {\rm Min} (v_F t,\xi)$.
 In particular, one must have $L < v_F t$, or $\Omega \sim 1/t < v_F q$.
\bibitem{signfootnote} To lowest order in perturbation theory in $\Gamma_t$
 one finds $c_{d-1}<0$, see Ref.\ \onlinecite{chi_s}. In that reference,
 various mechanisms have been discussed that may lead to $c_{d-1}>0$ for
 realistic values of the interaction strength. In what follows we dicuss
 only systems whose parameter values are such that $c_{d-1}>0$, since else
 the continuous ferromagnetic phase transition we are interested in does not 
 exist.
\bibitem{crossoverfootnote} For all dimensions $d>2$, where there is a 
 classical Heisenberg transition, there is still another crossover in the
 system, namely from the classical Gaussian region to the classical
 Heisenberg critical region. For $1<d<2$, where there is no long-range order 
 at any nonzero temperature, this is not the case. In either case,
 $\phi$ describes the leading low-temperature effect due to the relevance of 
 the temperature with respect to the quantum fixed point.
\end{references}
\end{document}